# Evidence of Higher Order Topology in Multilayer WTe$_2$ from Josephson Coupling through Anisotropic Hinge States

Yong-Bin Choi[1], Yingming Xie[2], Chui-Zhen Chen[2,†], Jin-Ho Park[1], Su-Beom Song[3], Jiho Yoon[4], Bum Joon Kim[1,5], Takashi Taniguchi[6], Kenji Watanabe[6], Hu-Jong Lee[1], Jong-Hwan Kim[3], Kin Chung Fong[7,*], Mazhar N. Ali[4,*], Kam Tuen Law[2,*] and Gil-Ho Lee[1,*]

[1]Department of Physics, Pohang University of Science and Technology, Pohang, Republic of Korea

[2]Department of Physics, Hong Kong University of Science and Technology, Clear Water Bay, Hong Kong, China

[3]Department of Materials Science and Engineering, Pohang University of Science and Technology, Pohang, Republic of Korea

[4]Max Plank Institute for Microstructure Physics, Halle (Saale), Germany

[5]Center for Artificial Low Dimensional Electronic Systems, Institute for Basic Science (IBS), Pohang, Republic of Korea

[6]Research Center for Functional Materials, National Institute for Materials Science, Tsukuba, Ibaraki, Japan;

[7]Raytheon BBN Technologies, Quantum Information Processing Group, Cambridge, MA, USA

†Current address: Institute for Advanced Study and School of Physical Science and Technology, Soochow University, Suzhou 215006, China

*Correspondence and requests for materials should be addressed to K.C.F (kc.fong@raytheon.com), M.N.A. (maz@berkeley.edu), K.T.L. (phlaw@ust.hk), or G.-H.L. (lghman@postech.ac.kr).

**The noncentrosymmetric Td-WTe$_2$, previously known as a type-II Weyl semimetal, is expected to have higher order topological phases with topologically protected, helical one-dimensional (1D) hinge states when their scarcely separated Weyl points get annihilated. However, the detection of these hinge states is difficult in the presence of the semimetallic behaviour of the bulk. Here, we spatially resolved the hinge states by analysing the magnetic field interference of supercurrent in Nb-WTe$_2$-Nb proximity Josephson junctions. The Josephson current along the *a*-axis of the WTe$_2$ crystal, but not along the *b*-axis, showed sharp enhancements at the edges of the junction; the amount of enhanced Josephson current was comparable to the upper limits of a single 1D conduction channel. Our experimental observations provide evidence of the higher order topological phase in WTe$_2$ and its corresponding anisotropic topological hinge states, in good agreement with theoretical calculations. Our work paves the way for hinge transport studies on topological semimetals in superconducting heterostructures, including their topological superconductivity.**

Three-dimensional (3D) Dirac and Weyl semimetals (WSMs) are materials which support massless fermionic quasipartcles[1]. WTe$_2$ represented the first example of a new kind of type-II WSM that hosts emergent quasiparticles of Lorentz-violating Weyl fermions, which were forbidden as a fundamental particle in nature and have therefore long been overlooked in quantum field theory[2,3]. Unique electronic transport signatures, including the intrinsic anomalous Hall effect[4-6], magnetic breakdown and Klein tunnelling effect[7], Landau level spectrum collapse[8] and the asymmetric Josephson effect[9], have been subjected to fundamental and application-based studies. WTe$_2$ is naturally close to a superconducting phase; it was shown to have low-pressure induced superconductivity while its sister phase, MoTe$_2$, was shown to be the only intrinsically superconducting WSM at 0.3 K with an extreme pressure-driven critical temperature enhancement[10,11]. Topological boundary states emerge at the surface of bulk WSMs in the form of Fermi arcs that connect projected Weyl points with different chirality in momentum space. However, the small separation of Weyl points in momentum space[2] makes Fermi arcs of bulk WTe$_2$ difficult to resolve by angle-resolved photoemission spectroscopy (ARPES)[12,13]. In the monolayer limit of WTe$_2$, a two-dimensional (2D) topological insulating phase was predicted to accompany one-dimensional (1D) quantum spin Hall edge states due to large spin-orbit coupling[14], and experimentally verified by electrical transport[15,16], ARPES and scanning tunnelling microscopy (STM)[17-19].

Recently, it was proposed that bulk $MoTe_2$ and $WTe_2$ in Td structures are higher order topological insulators (HOTIs), and that large surface states arising from gapped fourfold Dirac surfaces would be present[20]. Their signatures were actually observed previously but not understood; large surface states not attributable to the Fermi arcs arising from the Weyl points were seen by ARPES and STM in bulk $WTe_2$ and $MoTe_2$ [21,22]. In addition, $WTe_2$ was proposed to have topologically protected conducting helical hinge states[20], which would coexist with bulk conduction states and be difficult to resolve by transport measurements. However, these hinge states are difficult to observe in transport experiments due to the large contribution from the bulk states. In this work, we couple a $WTe_2$ thin film to two superconducting leads formed by Nb and measure the critical superconducting current of this Nb-$WTe_2$-Nb Josephson junction as a function of applied magnetic field. By analysing the Fraunhofer pattern resulted from the interference of the superconducting currents, we found evidence of conducting channels localised at the edges of the sample, consistent with the presence of topological hinge states due to the higher order topology of $WTe_2$. We further found that the hinge states are highly anisotropic; they are localised on edges parallel to the *a*-axis of the crystal while the hinge states along the *b*-axis are delocalised and merge with the bulk states. Our claims of the presence of hinge states are supported by theoretical calculations that explain the observed current density profile and Fraunhofer patterns.

Firstly, we describe the angle-dependent electrical transport in multilayer $WTe_2$. High-quality single crystals of $WTe_2$ used in this report were grown by the flux method[23,24]. There was no intrinsic chemical doping of $WTe_2$ crystals as confirmed by analysis of Shubnikov-de Haas oscillation in the bulk crystal or classical Hall measurements of the exfoliated multilayer $WTe_2$ (Fig. S1). Unlike many other transition metal dichalcogenides, such as $MoS_2$, $MoSe_2$ and $MoTe_2$, which have hexagonal planar structures, $WTe_2$ crystallises in the Td structure (orthorhombic SG# 31 Pmn21), wherein the hexagonal plane is distorted by the formation of tungsten chains running along the *a*-axis (Fig. 1a). This leads to the WSM bulk state and the 2D topological insulating state in the monolayer limit[14]. To investigate the in-plane electrical anisotropy of the crystal, we fabricated a multi-terminal circular $WTe_2$ device covered with hexagonal boron nitride (Fig. 1b). Angular dependence of the in-plane resistance, $R(\theta)$, within a single device was measured with a four-probe configuration to avoid contact resistance. Here, we chose the pair of voltage probes nearest to the pair of current electrodes, as shown in Fig.

1b, where $\theta$ denotes the angle between the direction of the resistance measurement (white line) and the *a*-axis of the WTe₂ crystal (red line). As expected from the crystalline anisotropy, the minimum resistance was observed along the *a*-axis ($\theta = 0$) and the maximum resistance was observed along the *b*-axis ($\theta = 90°$) (Fig. 1c). The ratio of maximum to minimum resistance reached 2.7 at 50 K and dropped to 2.3 at 4.2 K (inset of Fig. 1c). Other manifestations of such anisotropic transport of WTe₂ have been studied in quantum oscillations[25] and the chiral anomaly[26]. As the anisotropic crystallographic axes are also reflected in the symmetry of lattice vibrations, polarisation-resolved Raman spectroscopy[6,27,28] is used to determine the crystal axis for the following devices in an accurate and non-invasive manner. The A1 Raman mode at around 165 cm$^{-1}$ shows a characteristic two-fold pattern in the parallel polarisation configuration of laser excitation and detection (Fig. 1d). The *a*-axis (*b*-axis) of the crystal can be determined unambiguously based on the angle with maximum (minimum) intensity, as reported previously (see Figs. S2 and S3 for more data of Raman spectroscopy).

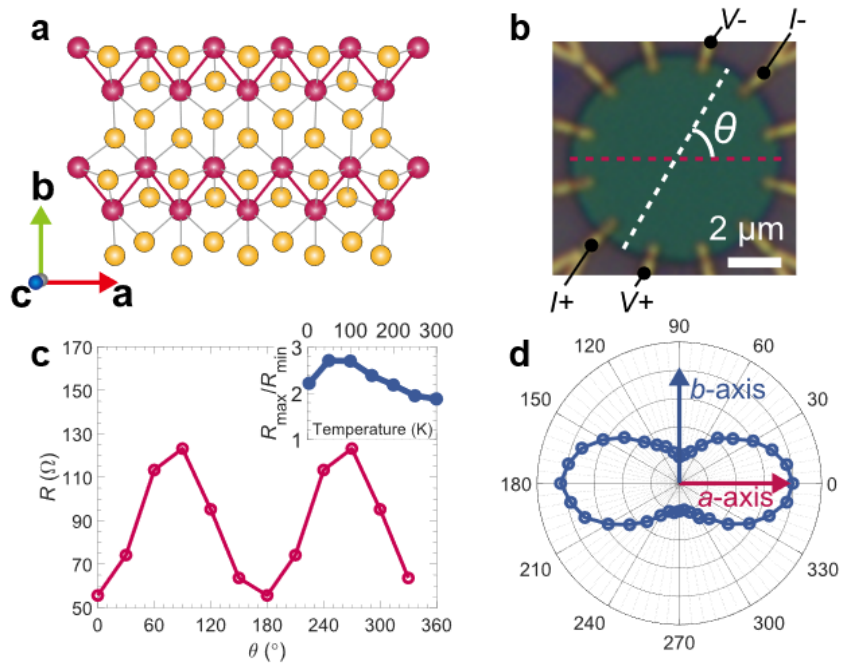

**Figure 1 | Anisotropy of WTe₂ crystal. a**, Schematic of the atomic structure of WTe₂ in the *a*-*b* plane. Red (yellow) balls represent the W (Te) atoms. A chain of W atoms forms along the *a*-axis. **b**, Optical micrograph of a 15-nm thick WTe₂ flake etched into a circular disk shape with multiple gold contact electrodes. $\theta$ is the angle between the measurement direction (white line) and the *a*-axis of the crystal (red line). **c**, Angular dependence of the resistance of the circular

device at 4.2 K. Resistance along the *a*-axis (*b*-axis) shows the smallest (largest) value. (Inset) Temperature dependence of the ratio of the maximum to minimum resistance. **d**, Polarisation angle dependence of the A1 phonon mode at ~165 cm$^{-1}$. The Raman signal shows the maximum (minimum) intensity along the *a*-axis (*b*-axis).

Now, we discuss the proximity Josephson junction created with the multilayer WTe$_2$, where WTe$_2$ is placed between two Nb superconducting electrodes as depicted in Fig. 2a. When superconducting contacts are electrically transparent, the superconducting order parameters of Nb electrodes can extend into WTe$_2$ and eventually lead to Josephson coupling, as mediated by WTe$_2$. We fabricated eight different Josephson junctions of length $L$ between 60 and 230 nm, width $W$ between 1.8 and 5.2 μm and WTe$_2$ thickness $t$ between 13 and 27 nm. These devices were divided into two groups with different crystal orientations, the *a*- and *b*-axes, aligned to the current direction (see Methods for fabrication processes). We will focus on two representative devices, Dev. A of ($L,W,t$) = (190 nm, 2.3 μm, 13.2 nm) with current flow along the *a*-axis and Dev. B of ($L,W,t$) = (100 nm, 3.3 μm, 25.0 nm) with current flow along the *b*-axis (see Table S1 for a summary of other devices). Figure 2c shows typical current-voltage (*I-V*) characteristics of Dev. A at various temperatures. While sweeping the bias current from zero to a positive value, the junction voltage switched from zero to a finite value at the Josephson critical current, $I_c$ = 0.7 μA. The retrapping current at which the voltage switched from a finite value back to zero was 0.4 μA, which was smaller than $I_c$ presumably due to self-heating of the junction[29]. The $I_c$ decreased monotonically with increasing temperature $T$ and eventually vanished around $T$ = 1.0 K, above which the *I-V* curve became linear with a slope corresponding to the normal resistance, $R_N$. As $I_c$ scales with cross-sectional area $A = Wt$, while $R_N$ scales with $1/A$, the $I_cR_N$ product became independent of $A$ such that it characterised the strength of Josephson coupling, taking account of the junction geometry. The measured $I_cR_N$ product of Dev. A (22.4 μV) was about seven times larger than that of Dev. B (3.0 μV). Anisotropy of the $I_cR_N$ product along the *a*- and *b*-axes was consistently observed for other six devices. The *T*-dependence of $I_c$ in Fig. 2d also showed more robust Josephson coupling of Dev. A than Dev. B.

Unlike the insulating bulk states of topological insulators, bulk states of WSMs, like WTe$_2$, are conducting, making it challenging to resolve the proposed topological hinge states.

Our approach was to spatially resolve how the Josephson current flows through the Nb-WTe$_2$-Nb Josephson junction by analysing its magnetic field response. When the magnetic field $B$ is threading the junction area, the Aharonov-Bohm effect leads to a macroscopic quantum phase difference between the superconductors and modulates $I_c$ according to the dc Josephson relationship. If a sinusoidal current-phase relation is assumed, the Josephson current interference pattern is determined by the maximum value of the phase-sensitive integration as $I_c(B) = \left|\int_{-\infty}^{\infty} J(x)e^{i\alpha x}dx\right|$, with Josephson current density $J(x)$, where $x$ is the real-space coordinate along the junction width direction, the magnetic field dependent parameter $\alpha = 2\pi BL_{\text{eff}}/\Phi_0$, the effective junction length $L_{\text{eff}} = L + 2L'$ taking into account the magnetic flux penetrating each superconducting electrode by $L'$, the magnetic flux quantum $\Phi_0 = h/2e$, Plank's constant $h$, and electron charge $e$. As $I_c(B)$ can be viewed as the magnitude of the Fourier transform of $J(x)$, we can extract $J(x)$ from the inverse Fourier transform (IFT) of the experimentally measured $I_c(B)$ (Fig. S4) and visualise the current flow density in the junction. This technique has been used for visualising edge-dominant transport of HgTe-based 2D quantum spin Hall insulators (QSHIs)[30,31] and surface-dominant transport of Bi-based 3D topological insulators[32]. A related technique with SQUID loop recently provided evidence of hinge states in Bi nanowires[33].

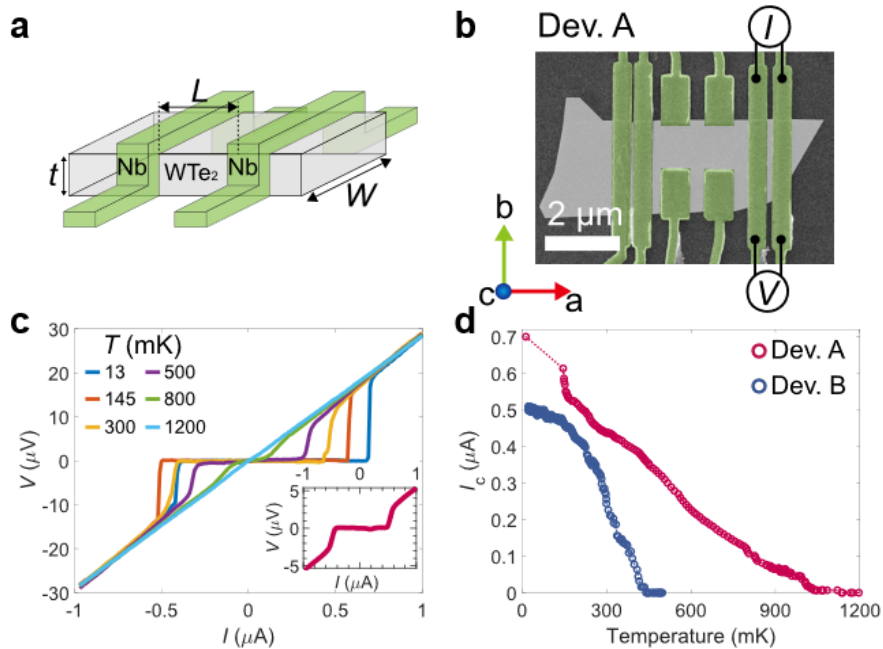

**Figure 2 | Characteristics of WTe$_2$-based Josephson junction. a**, Schematic of a WTe$_2$

Josephson junction. $W$, $L$, and $t$ represent the junction width, junction length, and thickness of the multilayer WTe$_2$ flake, respectively. **b**, False-colour scanning electron micrograph of Dev. A showing the measurement configuration. **c**, Typical current-voltage (*I-V*) characteristics of Dev. A measured by sweeping the bias current from negative to positive at different temperatures. The lower right inset shows the *I-V* characteristics of Dev. B at 22 mK without an external magnetic field. **d**, Temperature dependence of the Josephson critical current, $I_c$, for Dev. A and Dev. B.

Figures 3a and 3d show schematics of Dev. A and Dev. B, respectively, where Nb superconducting electrodes contact the top and both sides of the WTe$_2$ flake. The topological hinge states are expected to reside on opposite sides of the hinges of the WTe$_2$ flake (see Fig. S5 for more details). The Fraunhofer pattern of Dev. A is shown in Fig. 3b. The period of the $I_c$ oscillation, $\Delta B = 14.7$ G, corresponded well to the flux quantum threading of an effective junction area of $WL_{\text{eff}}$ with $L' \approx 210$ nm, ruling out the possibility of accidental micro-shorts between the Nb electrodes. $L'$ corresponds to the half width of the Nb electrodes, which can be understood by the magnetic flux fully penetrating the Nb electrode. The non-vanishing behaviour of $I_c(B)$ up to $B = 200$ G, shown in Fig. 3b, clearly deviated from the single-slit Fraunhofer pattern (black line in Fig. 3b) expected for a spatially uniform Josephson current density, $J(x)$ = constant. Qualitatively, the non-vanishing behaviour of $I_c$ resembles a SQUID interference pattern for a delta function-like Josephson current density at the junction edges. The $J(x)$ extracted by IFT of $I_c(B)$ for Dev. A (Fig. 3c) clearly showed enhancement of Josephson current density at both edges of the junction, with the baseline arising from the spatially uniform bulk contribution (black dotted line). This was consistent with the expectation of hinge states along the *a*-axis of the WTe$_2$ crystal. The width of the Josephson current-carrying hinge states was determined to be around 100 nm using a Gaussian line shape (red dotted line in Fig. 3c). Importantly, we performed a control experiment by fabricating a similar Josephson junction on topologically trivial graphite, to exclude the possibility that the non-vanishing pattern was an artefact (see Fig. S6); as expected, there was no evidence of the non-vanishing pattern. In contrast to Dev. A, Dev. B (Fig. 3d) exhibited an interference pattern resembling the conventional single-slit Fraunhofer pattern, where oscillation amplitude decayed with $1/B$ and vanished above ~50 G (Fig. 3e). This was very similar to the results of our control experiment with graphite. The Josephson current density extracted by IFT showed a uniform current distribution, without any signatures of edge-enhanced transport within the limits of our experimental resolution (Fig. 3f). Six more devices with different crystal axes showed consistent interference patterns and current density distributions (Fig. S7 for *a*-axis and Fig. S8 for *b*-axis). The sharp difference in $J(x)$ between Dev. A and Dev. B will be discussed in terms of the distinct shape of the spatial wave functions of topological hinge states along the different crystal axes.

We theoretically investigated the hinge states, and their Fraunhofer patterns, by adopting the model Hamiltonian reported previously[20,34], which describes the higher order topological properties of WTe$_2$ (see Methods for details). Our model showed highly anisotropic band structures (insets of Figs. 3h and 3j), with the helical hinge states showing markedly different localisation along the *a*- versus *b*-axis. The wave function of the hinge states along the *a*-axis was localised at the edges of the junction (Fig. 3h), matching the current density profile of Dev. A (Fig. 3c). When the chemical potential was set to introduce a bulk contribution, the calculated interference pattern in Fig. 3g, taking into account both hinge and bulk states, showed a close resemblance to the interference shape measured in Dev. A (Fig. 3b), especially for the non-vanishing behaviour of $I_c$ even at high $B$. In contrast, the hinge states along the *b*-axis were strongly merged into bulk states in momentum space (inset of Fig. 3j). Consequently, the wave function was well delocalised over the whole junction in real space, as plotted in Fig. 3j, which matched the uniform current density profile of Dev. B (Fig. 3f). This resulted in a single-slit Fraunhofer pattern, as shown in Fig. 3i, which was similar to the interference pattern measured for Dev. B (Fig. 3e). The observed Josephson current mediated by a single edge of the junction ($I_{J,\text{side}}$ = 22 nA for the left edge and 50 nA for the right edge, visible as the shaded areas in Fig. 3c) was comparable to and, importantly, less than the maximum theoretical value of $I_{J,h}$ = 140 nA for a single hinge state. We arrived at this maximum by considering a short ballistic junction limit ($eI_{J,h}R_{N,h} = \pi\Delta_{Nb}$), where $R_{N,h} = h/e^2$ is the normal resistance for a single hinge state and $\Delta_{Nb} = 1.763 k_B T_{c,Nb}$ is the BCS superconducting gap of the Nb electrode with $T_{c,Nb}$ = 7.5 K. $I_{J,\text{side}}$ being smaller than $I_{J,h}$ was consistent with the theory that there was only one hinge state per side of the WTe$_2$ flake, as depicted in Figs. 3a and 3d. Similar quantitative issues have been previously discussed in other HOTI candidates[33,35].

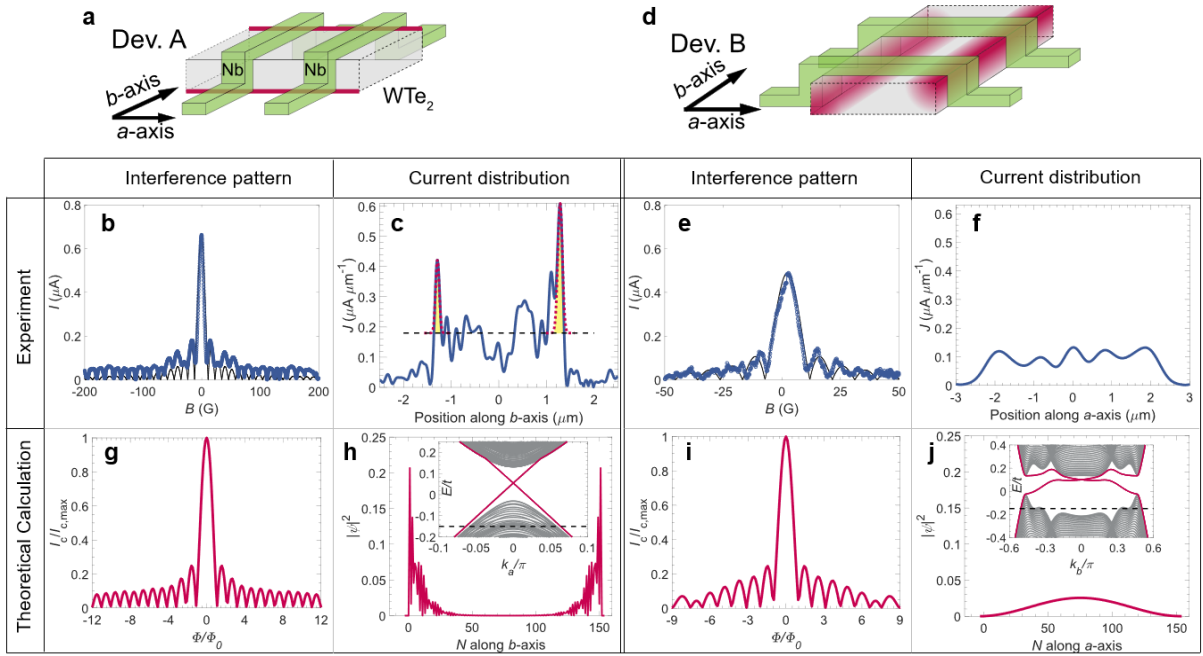

**Figure 3 | Interference pattern for the WTe$_2$ Josephson junction and extracted current profile. a**, **d**, Schematics of WTe$_2$ Josephson junctions with well-localised hinge states along the *a*-axis, represented by thick red lines (**a**, Dev. A), and with delocalised hinge states along the *b*-axis, conceptually represented by blur red lines (**d**, Dev. B) **b**, **e**, Magnetic field interference patterns of the critical current for Dev. A (**b**) and Dev. B (**e**). Black lines represent a single-slit Fraunhofer pattern of $|\sin(\pi L_{\mathrm{eff}} W B/\Phi_0)/(\pi L_{\mathrm{eff}} W B/\Phi_0)|$. **c**, **f**, Extracted spatial distribution of the Josephson current density *J* for Dev. A (**c**) and Dev. B (**f**). Red dotted lines in **c** represent Gaussian fitting for the edge current enhancements on each side of the junction, with the baseline arising from the bulk contribution (black dotted line). **g**, **i**, Theoretically calculated interference patterns of the critical current along the *a*-axis (**g**) and *b*-axis (**i**). **h**, **j**, Theoretically calculated wave functions of the hinge states along the *a*-axis (**h**) and *b*-axis (**j**). (Insets) The energy spectra of multilayer WTe$_2$ along the *a*-axis (**h**) and *b*-axis (**j**). The red lines in the energy spectra represent the helical hinge states and the grey lines represent the bulk states.

We note that there is an alternative, but less likely, explanation for the observation that stems from the QSHI phase of monolayer WTe$_2$. When monolayers of WTe$_2$ with 1D quantum spin Hall edge states are vertically stacked with finite interlayer interaction, 1D topological edge states gradually turn into the Fermi arc surface state of WSM. In this few-layer limit between the bulk and monolayer, there may exist edge states due to quantisation of the *c*-axis momentum in the Fermi arc states. However, this is unlikely for three reasons. Firstly, we were closer to the bulk limit in our experiments (18 – 38 layers) than the monolayer limit. Secondly, many experiments have been carried out on few-layer (2 – 10 layers) WTe$_2$, and edge states have been seen strictly in the monolayer[15-19]. Finally, the Fermi arcs from the Weyl points in WTe$_2$ are extremely short and difficult to resolve by ARPES, such that edge states originating from quantisation of Fermi arcs show greater mixing with the bulk states.

By investigating the transport of multilayer WTe$_2$ in both normal and superconducting regimes in combination with model Hamiltonian calculations, we provided evidence of higher order topology of WTe$_2$ via proximity-induced superconductivity in anisotropic hinge states. Several theoretical studies on the effect of superconducting proximity on WSMs showed that the Andreev reflection with bulk states is supressed due to conservation of chirality[36], while the topological boundary state interacts well with superconductivity[37,38]. Such selective coupling of bulk and boundary states to superconductors makes superconducting heterostructures an ideal platform for investigating topological semimetals that have mixed bulk and topological boundary contributions, particularly HOTI systems.

**Methods**

**Device fabrication.** WTe$_2$ single crystal was mechanically exfoliated into WTe$_2$ thin film of thickness about 10 to 20 nm and placed on Si substrate covered with 300-nm thick insulating SiO$_2$. Thickness of WTe$_2$ flakes was confirmed by atomic force microscopy after the electrical measurement. As bulk single crystal has a needle shape elongated along the *a*-axis, crystal axis could be tracked during the exfoliation process. Polarised Raman spectroscopy was used to confirm the crystal axis direction after the electrical measurement. To eliminate possibly contaminated topmost layers of WTe$_2$ flake, *in-situ* Argon ion etching was conducted before the metal deposition for removing possibly oxidized WTe$_2$ surface layers due to ambient air. Ti/Au (5 nm/100 nm) or Ti/Nb/Au (5 nm/100 nm/5 nm) electrode were deposited for studying normal transport or superconducting transport, respectively. Ti layer improves adhesion of metal electrodes to WTe$_2$, and Au layer protects Nb superconducting electrode from oxidation. The chamber pressure during Ti evaporation was kept lower than $8 \times 10^{-8}$ Torr. Nb layer in Ar atmosphere was sputtered at slow rate of 8.33 nm/min to minimise the damage of WTe$_2$ flake by Ar plasma. In the whole fabrication process, top surface of WTe$_2$ thin film was exposed Poly-methyl methacrylate (PMMA) polymer only one time to minimise the degradation of WTe$_2$.

**Polarisation-resolved Raman spectroscopy.** Raman spectra are obtained with a confocal microscopy set-up at room temperature. The samples are excited with HeNe laser (632.8 nm) at normal incidence. Raman signal is collected in the backscattering configuration and analysed by a monochromator equipped with liquid nitrogen-cooled silicon CCD. Two linear polarisers in the parallel configuration are placed immediately after the laser and before the monochromator to define the polarisations of incident and scattered light, respectively. The crystal orientation relative to the polarisation is controlled with a half waveplate between a beam splitter and WTe$_2$ flakes.

**Effective model.** The Td-WTe$_2$ has been identified as a type-II WSM[2]. However, according to both ab initial calculation[13] and ARPES measurement[12], its Fermi arc state connecting Weyl points is very short. Observed large arc-like surface states in most previous works are determined to be topologically trivial, and Wang et al.[20] recently proposed that large arc-like state are actually the split and gapped Dirac cone surface states in HOTI phase. The HOTI phase in 1T′-WTe$_2$ or MoTe$_2$ (monoclinic SG# 11 P2$_1$/m) has been justified to be driven by the double band inversion at Gamma point[20,39]. From 1T′-WTe$_2$ to Td-WTe$_2$, not only the Weyl points are created, but also the huge arc-like surface states are inherited from original HOTI phase. Considering the small separation between the bulk Weyl points, the HOTI phase in Td-WTe$_2$ may be readily realised under some realistic conditions, such as the presence of small strain[20,40]. Since the 1T′- and Td-WTe$_2$ in HOTI phase exhibit the same topological surface state feature, we would not be able to distinguish them in this work. As we want to probe the topological feature of WTe$_2$, it is sufficient to use a model Hamiltonian with equivalent topological nature as[20,34],

$$H(\mathbf{k}) = H_{MN}(\mathbf{k}) + V_c + V_{so}(\mathbf{k}). \quad - \text{Eq.}(1)$$

Here, $H_{MN}(\mathbf{k}) = \left(m_1 + \sum_{j=a,b,c} v_j \cos(k_j) + m_2 \mu^x + m_3 \mu^z\right)\tau^z + \lambda_b \sin k_b \mu^y \tau^y + \lambda_c \sin k_c \tau^x$, $V_c = \gamma_x \mu^x + \gamma_z \mu^z$, $V_{so}(\mathbf{k}) = \beta_a \sin k_a \mu^y \tau^y \sigma^z$. The Pauli matrices $\mu, \tau$ operates on different orbital space, $\sigma$ operates on spin space. The Hamiltonian $H_{MN}(\mathbf{k})$ gives a bulk monopole node-line (MNL) semimetal[20], $V_c$ is a mass term that titles the bulk MNL,

$V_{so}(\boldsymbol{k})$ gaps the bulk MNL and gives the helical hinge states. The model Hamiltonian $H(\boldsymbol{k})$ exhibits hinge states and has similar higher order topological properties of WTe$_2$ or MoTe$_2$ [20,34]. In the main text Figs. 3g-j, we set $m_1 = -3t$, $m_2 = 0.3t$, $m_3 = 0.2t$, $v_a = 2t$, $v_b = 1.6t$, $v_c = t$, $\lambda_b = 0.1t$, $\lambda_c = t$, $\gamma_x = 0.4t, \gamma_z = -0.4t, \beta_a = 1.5t$. $t = 1$ is a unit of energy.

**Numerical calculation of Josephson current.** In the calculation, we assume multilayer WTe$_2$ underneath Nb electrodes are fully superconducting due to proximity effect and their pairing potentials are set by $\Delta e^{\pm i\phi_0/2}$, respectively. We model multilayer WTe$_2$ with the lattice model discretising from Hamiltonian $H(\boldsymbol{k})$ in Eq. (1) and the Josephson current was evaluated by recursive Green's function method[9] by setting the system with three layers of WTe$_2$, length to be 25 sites, width to be 150 sites, $k_B T = \Delta/200$, and $\Delta = 0.5$.

## Acknowledgements

Y.-B.C. and G.-H.L. were supported by Samsung Science and Technology Foundation under Project Number SSTF-BA1702-05 for device fabrications and low temperature measurements, and National Research Foundation of Korea (NRF) Grant funded by the Korean Government (No. 2016R1A5A1008184). J.K. and S.S. acknowledges the support from the National Research Foundation of Korea grant (No. 2017R1C1B2012729). B.J.K. acknowledges the support from the Institute for Basic Science (IBS-R014-A2). K.W. and T.T. acknowledge support from the Elemental Strategy Initiative conducted by the MEXT, Japan and the CREST (JPMJCR15F3), JST. C.-Z. C. and K.T.L acknowledge the support of The Croucher Foundation and HKRGC through C6026-16W, 16324216, 16307117 and 16309718.

## Author contributions

K.C.F., M.N.A., K.T.L., and G.-H.L. conceived and supervised the project. Y.-B.C. fabricated the samples. Y.-B.C., J.-H.P. and H.-J.L. performed transport experiments. J.Y. and M.N.A. provided the WTe$_2$ crystal, and T.T. and K.W. provided the hexagonal boron nitride crystal. S.-B.S., B.J.K., and J.-H.K. collected and analysed polarized Raman spectrum. Y.X., C.-Z.C, and K.T.L. performed and analysed theoretical calculations for band spectrums and Fraunhofer patterns. Y.-B.C., Y.X., K.C.F., M.N.A., K.T.L., and G.-H.L. wrote the paper with inputs from C.-Z.C., J.-H.P., S.-B.S., and J.-H.K.

# Supplementary Information of
# Evidence of Higher Order Topology in Multilayer WTe2 from Josephson Coupling through Anisotropic Hinge States

Yong-Bin Choi[1], Yingming Xie[2], Chui-Zhen Chen[2,†], Jin-Ho Park[1], Su-Beom Song[3], Jiho Yoon[4], Bum Joon Kim[1,5], Takashi Taniguchi[6], Kenji Watanabe[6], Hu-Jong Lee[1], Jong-Hwan Kim[3], Kin Chung Fong[7,*], Mazhar N. Ali[4,*], Kam Tuen Law[2,*] and Gil-Ho Lee[1,*]

[1]Department of Physics, Pohang University of Science and Technology, Pohang, Republic of Korea

[2]Department of Physics, Hong Kong University of Science and Technology, Clear Water Bay, Hong Kong, China

[3]Department of Materials Science and Engineering, Pohang University of Science and Technology, Pohang, Republic of Korea

[4]Max Plank Institute for Microstructure Physics, Halle (Saale), Germany

[5]Center for Artificial Low Dimensional Electronic Systems, Institute for Basic Science (IBS), Pohang, Republic of Korea

[6]Research Center for Functional Materials, National Institute for Materials Science, Tsukuba, Ibaraki, Japan;

[7]Raytheon BBN Technologies, Quantum Information Processing Group, Cambridge, MA, USA

†Current address: Institute for Advanced Study and School of Physical Science and Technology, Soochow University, Suzhou 215006, China

*Correspondence and requests for materials should be addressed to K.C.F (kc.fong@raytheon.com), M.N.A. (maz@berkeley.edu), K.T.L. (phlaw@ust.hk), or G.-H.L. (lghman@postech.ac.kr).

## S1. Magneto-transport on bulk and thin-flake WTe$_2$

Figure S1a show the magnetoresistance measurement of bulk WTe$_2$ crystal. WTe$_2$ was observes large magnetoresistance (MR)[1], here we also observe large MR up to 10,000 % at the perpendicular magnetic field, MR is defined as $(R_{xx}(B) - R_0)/R_0$, $R_0 = R_{xx}(B=0)$. Nonlinear Hall resistance $R_{xy}(B)$ was observed because both electron and hole carrier type contribute Hall transport. Using classical two-band model for resistance, we assumed two carrier density and mobility[2,3].

$$R_{xy} = \frac{B[(n_h \mu_h^2 - n_e \mu_e^2) + (n_h - n_e) \mu_h \mu_e B^2]}{e[(n_h \mu_h + n_e \mu_e)^2 + (n_h - n_e)\mu_h^2 \mu_e^2 B^2]}$$

Electron carrier density and hole carrier density similar, but electron carrier density slightly larger than hole carrier density ($n_e = 4.08 \times 10^{25}\ m^{-3}, n_h = 3.89 \times 10^{25}\ m^{-3}$). From the classical two band model, electron and hole carrier mobility was calculated by $\mu_e = 2000 \times 10^{-4}\ m^2/V \cdot s, \mu_h = 2040 \times 10^{-4}\ m^2/V \cdot s$, respectively.

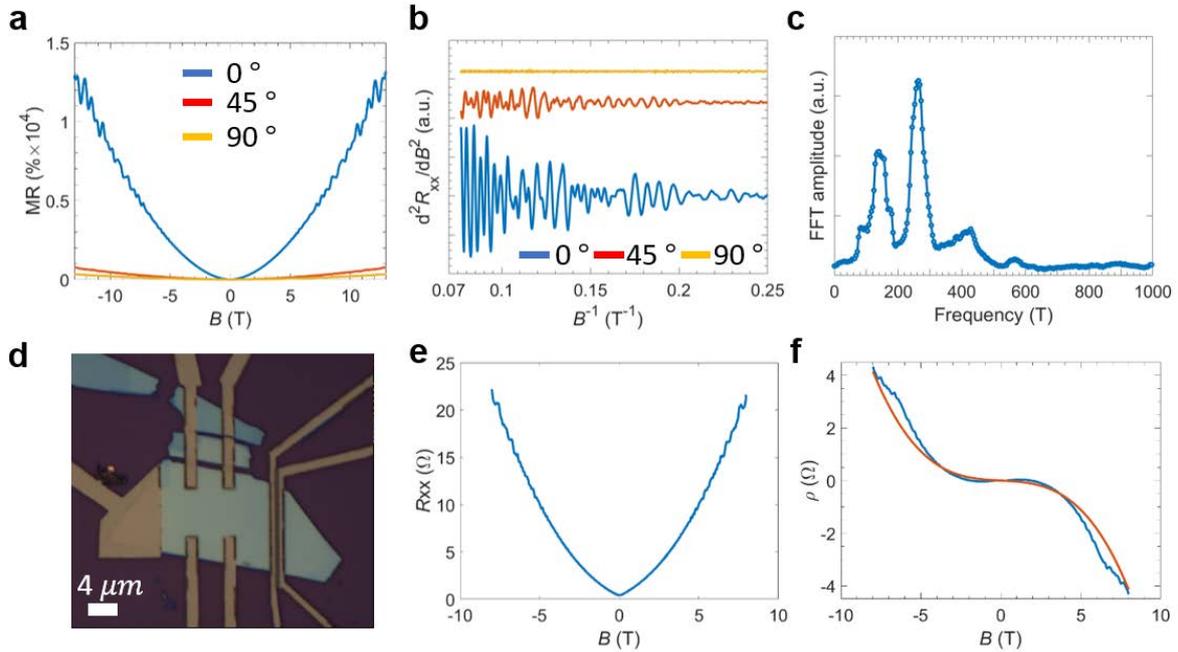

**Figure S1 | Magneto transport measurements on bulk and thin-film WTe$_2$. a,** Magnetoresistance (MR) with magnetic field perpendicular (blue), in-plane (yellow), and 45 ° (red) to the *ab* crystal plane. **b,** SdH oscillation as a function of inverse magnetic field perpendicular (blue), in-plane (yellow), and 45 ° (red) to the *ab* crystal plane. **c,** Fast Fourier transform of oscillatory magnetoresistance at 0 °. **d,** Optical micrograph of Hall geometric

WTe$_2$ device. **e,** Measured R$_{xx}$ of Hall geometric device at 290 mK. **f,** Measured R$_{xy}$ (blue) of Hall geometric device with classical two band model fitting (red).

## S2. Crystal axis identification from polarization-resolved Raman spectrum

Polarization-resolved Raman spectroscopy is utilized as an accurate and non-invasive characterization tool to determine the anisotropic crystal orientation. The polarizations of incident and scattered light are set in the parallel configuration while the crystal orientation of WTe$_2$ flakes relative the polarization is rotated by a half waveplate as described in methods section of the main text. The crystal axis can be determined from polarization-angle dependence of the Raman intensity following the previous works[6,21,22]. Figure S2a, d, and g (Figure S3. a, and b) shows micrograph of a WTe$_2$ flake of Dev. A1, Dev. A2, and Dev. A3 (Dev. B1, and Dev. B2), respectively. The polarization direction ($\theta$, red arrow) is varied from 0 to 360 degree with respect to the vertical direction (black solid line) of the micrograph. The Colour-coded plot of Raman intensity are shown in Fig. S2. b, e, and h (Figure S3. c, and d). The Raman signals originate from the A$_1$ modes of Td-WTe$_2$ crystals which all show polarization-angle dependence. For example, the Raman modes around at 165 cm$^{-1}$ and 213 cm$^{-1}$ show characteristic two-fold patterns (Figure S2.c, f, and i), (Figure S3. e, and f) in their polarization angle dependence. Intensity maximum of ~ 165 cm$^{-1}$ and 213 cm$^{-1}$ Raman modes appear when the polarization aligns respectively with the *a*-axis and *b*-axis, which is previously established based on high-resolution atomic force microscopy and Raman tensor analysis[6,21,22].

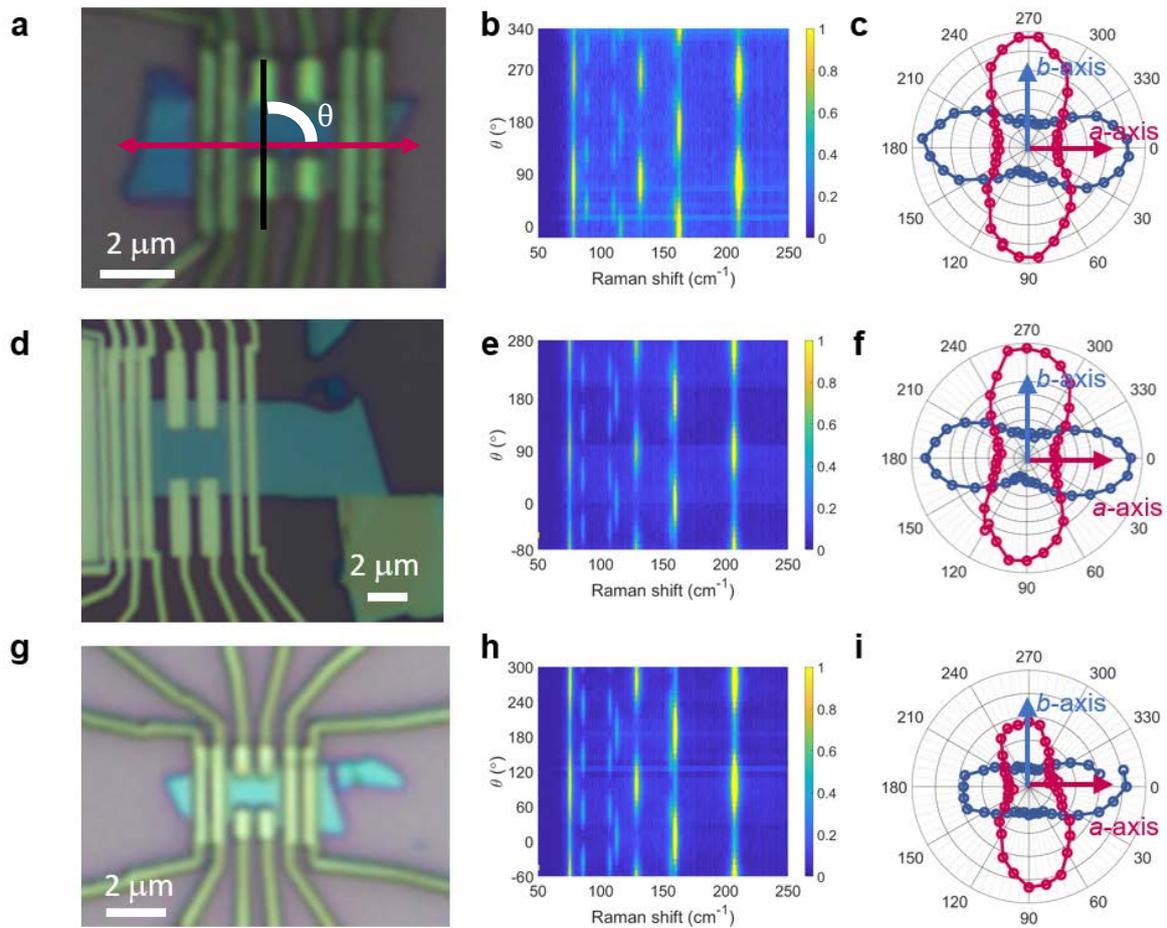

**Figure S2 | Polarized Raman spectroscopy for determining crystal axis for Dev. A1, A2, and A3. a, d, g** Optical micrograph of WTe$_2$ devices for Dev. A1 (**a**), Dev. A2 (**d**), and Dev. A3 (**g**), respectively. **b, e, h,** Colour-coded plot of Raman intensity as a function of relative crystal angle and relative Raman shift for Dev. A1 (**b**), Dev. A2 (**e**), and Dev. A3 (**h**), respectively. **c, f, i** Polar plots of Raman intensity with a function of the polarization angle. Raman modes at ~ 165 (blue circles and line) and 213 cm$^{-1}$ (red circles and line) show characteristic two-fold patterns where intensity maximums align to *a*-axis and *b*-axis of WTe$_2$ crystals for Dev. A1 (**c**), Dev. A2 (**f**), and Dev. A3 (**i**), respectively.

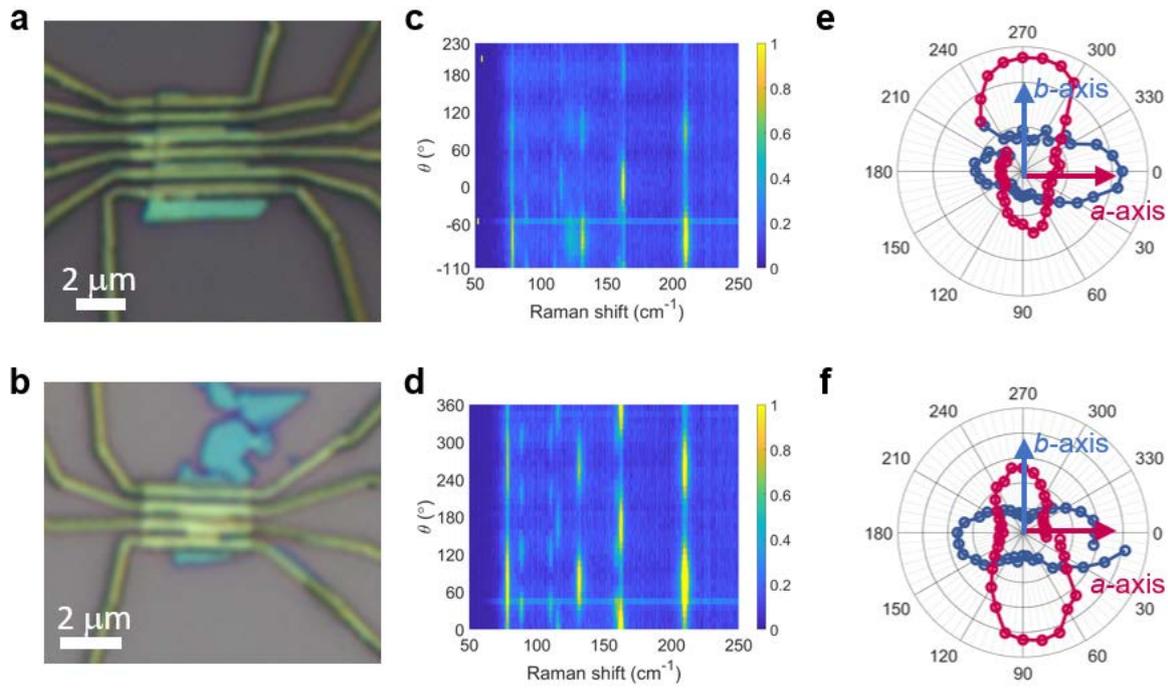

**Figure S3 | Polarized Raman spectroscopy for determining crystal axis for Dev. B1 and B2. a, b** Optical micrograph of WTe$_2$ devices for Dev. B1 (**a**), and Dev. B2 (**b**), respectively. **c, d,** Colour-coded plot of Raman intensity as a function of relative crystal angle and relative Raman shift for Dev. B1 (**c**), and Dev. B2 (**d**), respectively. **e, f,** Polar plots of Raman intensity with a function of the polarization angle. Raman modes at ~ 165 (blue circles and line) and 213 cm$^{-1}$ (red circles and line) show characteristic two-fold patterns where intensity maximums align to *a*-axis and *b*-axis of WTe$_2$ crystals for Dev. B1 (**e**), and Dev. B2 (**f**), respectively.

## S3. Summary of WTe$_2$ Josephson junction series of Dev. A, and Dev. B

We fabricated more WTe$_2$ Josephson junction device along *a*-axis and *b*-axis of the WTe$_2$ crystal of Dev. A1 to Dev. A4 and Dev. B1 to Dev.B4 and summarized in Table S1. Here, Dev. A1 and Dev. B1 represent Dev. A and Dev. B in the main text, respectively. We also observed non-vanishing interference pattern for series of Dev. A (Fig. S7) and conventional single slit interference pattern for series of Dev. B (Fig. S8)

| Name | | $I_C$ (μA) | $R_N$ (Ω) | $I_C R_N$ (μV) | Length (nm) | Width (μm) | Thickness (nm) | $R_N \times W \times T / L$ (μΩ·m) |
|---|---|---|---|---|---|---|---|---|
| Device along a-axis | Dev. A1 | 0.7 | 32 | 22.4 | 190 | 2.31 | 13.2 | 5.14 |
| | Dev. A2 | 0.9 | 7 | 6.3 | 230 | 5.18 | 22 | 3.47 |
| | Dev. A3 | 1.7 | 20 | 34 | 91 | 2.23 | 14 | 6.86 |
| | Dev. A4 | 2.5 | 3.5 | 8.8 | 110 | 4.30 | 27.2 | 3.72 |
| Device along b-axis | Dev. B1 | 0.5 | 6 | 3.0 | 100 | 3.33 | 25 | 5.00 |
| | Dev. B2 | 0.1 | 25 | 2.5 | 120 | 2.35 | 16.7 | 8.18 |
| | Dev. B3 | 0.08 | 30 | 2.4 | 130 | 1.82 | 14.6 | 6.13 |
| | Dev. B4 | 0.1 | 15 | 1.5 | 150 | 3.78 | 18.8 | 7.11 |

**Table S1 | Summary of devices**

## S4. Inverse Fourier transform technique

Josephson junction with a perpendicular magnetic field, the magnitude of the maximum critical current $I_c^{max}(B)$ is an absolute value of Josephson current $I_J(\phi)$ and the critical current of the junction modulated with external magnetic field and the modulation depending on Josephson current density in real space. For example, when uniform Josephson current density flow through the junction, the maximum critical current show single-slit Fraunhofer interference pattern. However, when Josephson current density flow only to the edge, the interference pattern shows SQUID interference pattern. In our experiment, we can assume non-vanishing Interference pattern caused by uniform and edge enhanced Josephson current density.

The total Josephson current through junction ($I_J$) and critical current are following the equations,

$$I_J(\phi) = \int_{-L/2}^{L/2} dx\, J(x) e^{i\phi(x)},  \quad \text{Eq. (1)}$$

$$I_c^{max} = |I_J(\phi)|,$$

Here, $J(x)$ is Josephson current distribution in real space, $\phi(x) = \phi_0 + 2\pi B L d_{\text{eff}} x / \Phi_0$, initial phase different $\phi_0$, effective distance $d_{\text{eff}}$, and magnetic flux quantum $\Phi_0 = h/2e$.

The equation of $I_J(\phi)$ is a Fourier transform of a Josephson current distribution in real space $J(x)$ Therefore, we can reconstruct Josephson current density in real space using inverse Fourier transform of Josephson current ($I_J$).

First, from the experimentally acquired interference pattern, extract critical current $I_C^{max}(B)$. For an ideal situation, Josephson current has only even current term, because odd part of Eq. (1) vanish from the integration. Thus, equation (1) becomes Fourier transform of even current density $I_J^E(\phi) = \int_{-L/2}^{L/2} dx\, J_E(x) \cos(\phi(x))$. Now then, from the critical current $I_c^{max} = |I_J(\phi)|$, we can recover the even Josephson current $I_J^E(\phi)$ by flipping the sign of every other lobe of the extracted $I_c^{max}$.

However, our situation is not an ideal such as non-zero resistance between two lobes. Thus, the odd part of Josephson current cannot be ignored, so total Josephson current should be considered not only even part but also odd part of Josephson current.

$$I_J(\phi) = I_E(\phi) + i I_O(\phi) = \int_{-L/2}^{L/2} dx\, [J_E(x)\, \cos(\phi(x)) + i J_O(x)\, \sin(\phi(x))],$$

Therefore, considering the odd part of Josephson current, the observed critical current become $I_c^{max} = |I_J(\phi)| = \sqrt{I_E^2(\phi) + I_O^2(\phi)}$. This equation means that when even Josephson current $I_E(\phi)$ is minimal, the odd Josephson current $I_o(\phi)$ is dominant. From this, $I_o(\phi)$ can be recovered (Fig. S4). By performing inverse Fourier transform, we can reconstruct Josephson current distribution in real space as[4]

$$J(x) = \left| \frac{1}{2\pi} \int_{-a/2}^{a/2} d\phi\, I_J(\phi) e^{-i\alpha x} \right|, \text{ with } \alpha = 2\pi B d_{\text{eff}} L / \Phi_0.$$

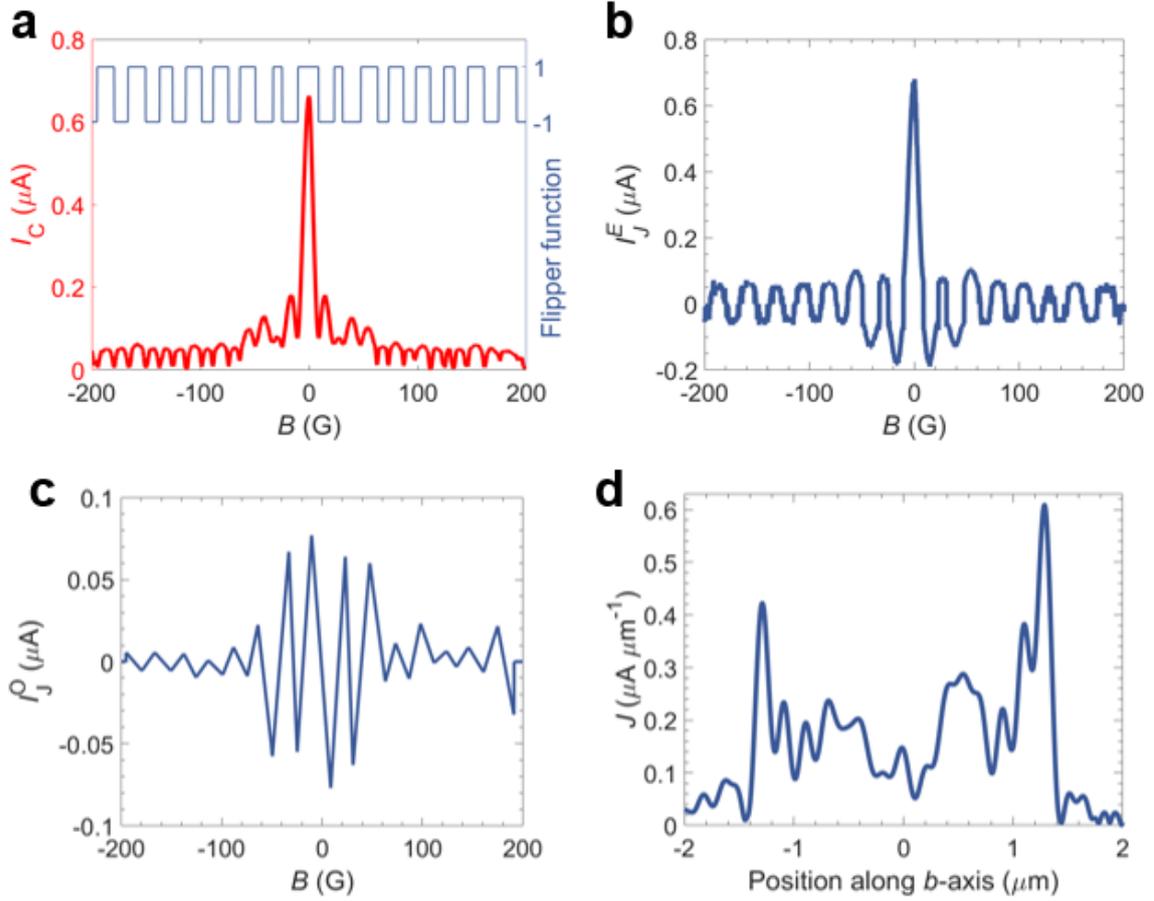

**Figure S4 | Inverse Fourier Transform of critical current. a,** (red)extracted critical current from observed Fraunhofer interference pattern and (blue)flipper function. **b,** The recovered even part of Josephson current $I_J^E(\phi)$. **c,** The recovered odd part of Josephson current $I_J^O(\phi)$. **d,** The reconstructed Josephson current density in real space along b-axis.

## S5. Bulk-Boundary correspondence

The wave-function of the hinge states given by $H(\mathbf{k})$ are shown in Fig. S5. It can be found that the hinge sates mainly locate at the opposite hinges of the 3D system along *a*-axis. There are some finite values of wave-function coming from the surfaces perpendicular to *a*-axis. In contrast, the surfaces parallel to *a*-axis are well gapped. This bulk-boundary correspondence can be understood from the two fold symmetry in $H(k)$, parity symmetry $P = \tau^z$ and mirror symmetry $M_a = i\sigma^y$, where $P\mathbf{k} \to -\mathbf{k}, M_a\mathbf{k} \to (-k_a, k_b, k_c)$. The location of the hinge states can be simply determined by the sign of mass $m_r$ at different surfaces[5], which is defined within the low energy surface Hamiltonian and gaps the Dirac surface states.

In the inversion symmetric higher order topological insulators (HOTIs), the mass changes sign under inversion: $m_{-r} \to -m_{-r}$. And according to the surface Dirac theory[5], the surface parallel to the mirror plane remains gapless in this scenario. The hinge states exists at where the two adjacent surface forms domain wall, namely the sign of mass changes. Thus, the localization feature of wave-function in Fig. S5 is consistent with boundary correspondence of the HOTI phase in presence of both inversion and mirror symmetry. More examples and detail analysis about the constraint of inversion and mirror symmetry in HOTI can be found in Ref.[5,6]. However, the inversion symmetry is broken for Td-WTe$_2$. In Ref.[7], they proposed Td-WTe$_2$ are very likely to be inversion-broken HOTI phase. In this case, there is no obvious constraint to forbid the mass terms even for the surface parallel to the mirror plane. However, we still attempt to believe that the hinge states are more localized for the hinges perpendicular to the mirror plane, considering the non-inversion symmetric HOTI in Td-WTe$_2$ is inherited from inversion symmetric HOTI in 1T′-WTe$_2$ [7].

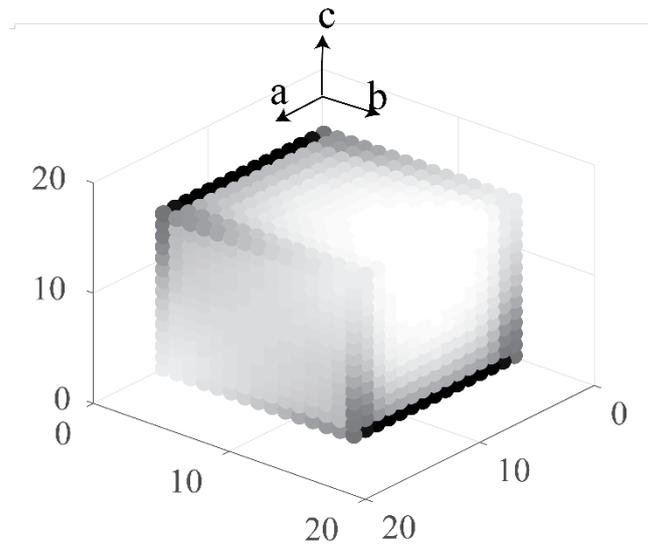

**Figure S5 |** Localized hinge modes of the minimal model of a HOTI with the same topology and symmetries as WTe$_2$ with open boundary conditions in all directions.

## S6. Graphite Josephson junction

We fabricate graphite-based Josephson junction as a control experiment of WTe$_2$ Josephson junction because graphite has isotropic band structure while WTe$_2$ has not and graphite-based Josephson junction has expected isotropic transport. The fabrication process completely same as a fabrication process of WTe$_2$ Josephson junction as described in main text. To reduce conduction of graphite, we choose thinner flake than WTe$_2$ flake (10~20 nm).

Here, we investigate two Josephson junctions in the one graphite device, GJJ-JJ1 of ($L,W,t$)=(250 nm, 7.1 μm, 5.8 nm) and GJJ-JJ2 of ($L,W,t$)=(200 nm, 7.1 μm, 5.8 nm) in Fig. S6a. The interference patterns were measured at 0.1 K (Figs. S6b and d). The critical current at field center is 3.5 μA for GJJ-JJ1 and 3.3 μA for GJJ-JJ2. The $I_cR_N$ products are 35 μV and 31 μV for GJJ-JJ1 and GJJ-JJ2, respectively, which are in the same order of magnitude of those of Dev. A and Dev. B. The oscillation period of Δ$B$=4.5 G (5.3 G) of GJJ-JJ1 (GJJ-JJ2) gives $L'$~200 (175) nm, which corresponds to the nearly half width of Nb electrodes. Reconstruction of uniform spatial distribution of Josephson current density (Figs. S6c, and e) inferred observed unconventional interference pattern of WTe$_2$ devices are not artefacts introduced during fabrication process.

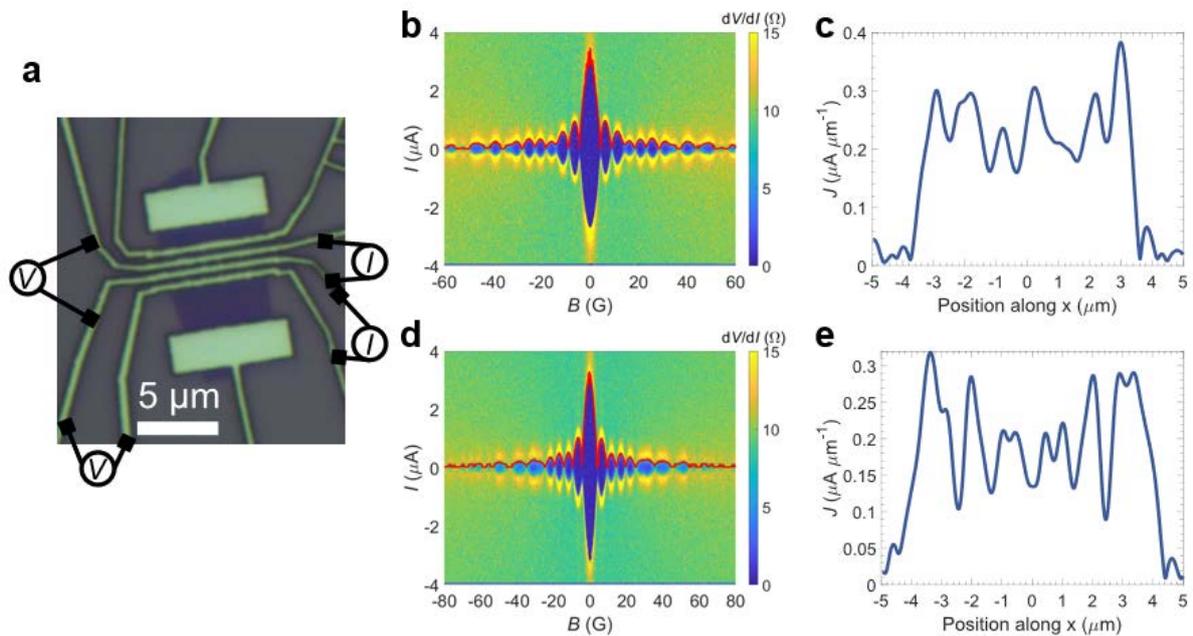

**Figure S6 | Control experiment device Graphite Josephson junction. a**, Optical micrograph of Graphite Josephson junction. **b**, **d,** Colour-coded plot of differential resistance d$V$/d$I$ as a function of bias current $I$ and perpendicular external magnetic field $B$ for JJ1 (**b**) and JJ2 (**d**). Red solid lines represent extracted Josephson critical current. **c, e**, Extracted spatial distribution of Josephson current density $J$ for GJJ-JJ1 (**c**) and GJJ-JJ2 (**e**).

## S7. Other WTe$_2$ Josephson junction devices along *a*-axis

Figure S7 show other Nb-WTe$_2$-Nb Josephson junction device of Dev. A2 (Figs. 7a, b, and c), Dev. A3 (Figs. S7d, e, and f), and Dev. A4 (Figs. S7g, h, and i). The fabrication process completely same as a described in main text. Here, Dev. A2, Dev. A3, and Dev. A4 was measured under 20 mK. Information of devices are summarized in Table S1. For Dev. A3 and Dev. A4 superconducting contact is Nb as same as Dev. A1 and series of Dev. B. Whereas, Dev. A2 used superconducting contact for Aluminum (Al). Therefore, the side lobes decayed over 70 G because of the small critical magnetic field (~100 G) of Al (Fig. S7b). The oscillation period of $\Delta B$=7.0 G gives $L'$~195 nm, which corresponds to the nearly half width of Al electrodes. Josephson current density in real space was reconstructed by using inverse Fourier transform and confirmed edge-enhanced transport. Gaussian fitting for each left(right) edge enhancement with width around 234(334) nm for Dev. A2. We observed SQUID-like interference pattern, the lobes dose not decay until 100 G for a Dev. A3. The oscillation period of $\Delta B$=14.7 G, and 8.85 G gives $L'$~170 nm, and 210 nm, which corresponds to the nearly half width of Nb electrodes for Dev A3 for Dev. A4, respectively. Using inverse Fourier transform, we check edge-enhanced transport and Gaussian fitting for each left(right) edge enhancement with width around 144(271) nm, and 372(575) nm for Dev A3, Dev A4, respectively. By considering a short ballistic junction limit ($eI_{J,h,Nb(Al)}R_{N,h} = \pi\Delta_{Nb(Al)}$) the maximum theoretical value of $I_{J,h,Nb(Al)} = 140(22)$ nA for a single hinge state, where $R_{N,h} = h/e^2$ is the normal resistance for a single hinge state and $\Delta_{Nb(Al)} = 1.763k_BT_{c,Nb(Al)}$ is the BCS superconducting gap of the Nb(Al) electrode with $T_{c,Nb(Al)} = 7.5(1.2)$ K. $I_{J,side}$ of Dev. A2, Dev. A3 and Dev. A4 is 8.8, 11 and 33 nA for left edge and 55, 65 and 143 nA for right edge, respectively, as visible as the shaded areas in Figs. S7c,f, and i. These values are comparable to or less than the maximum theoretical value $I_{J,h,Nb(Al)}$.

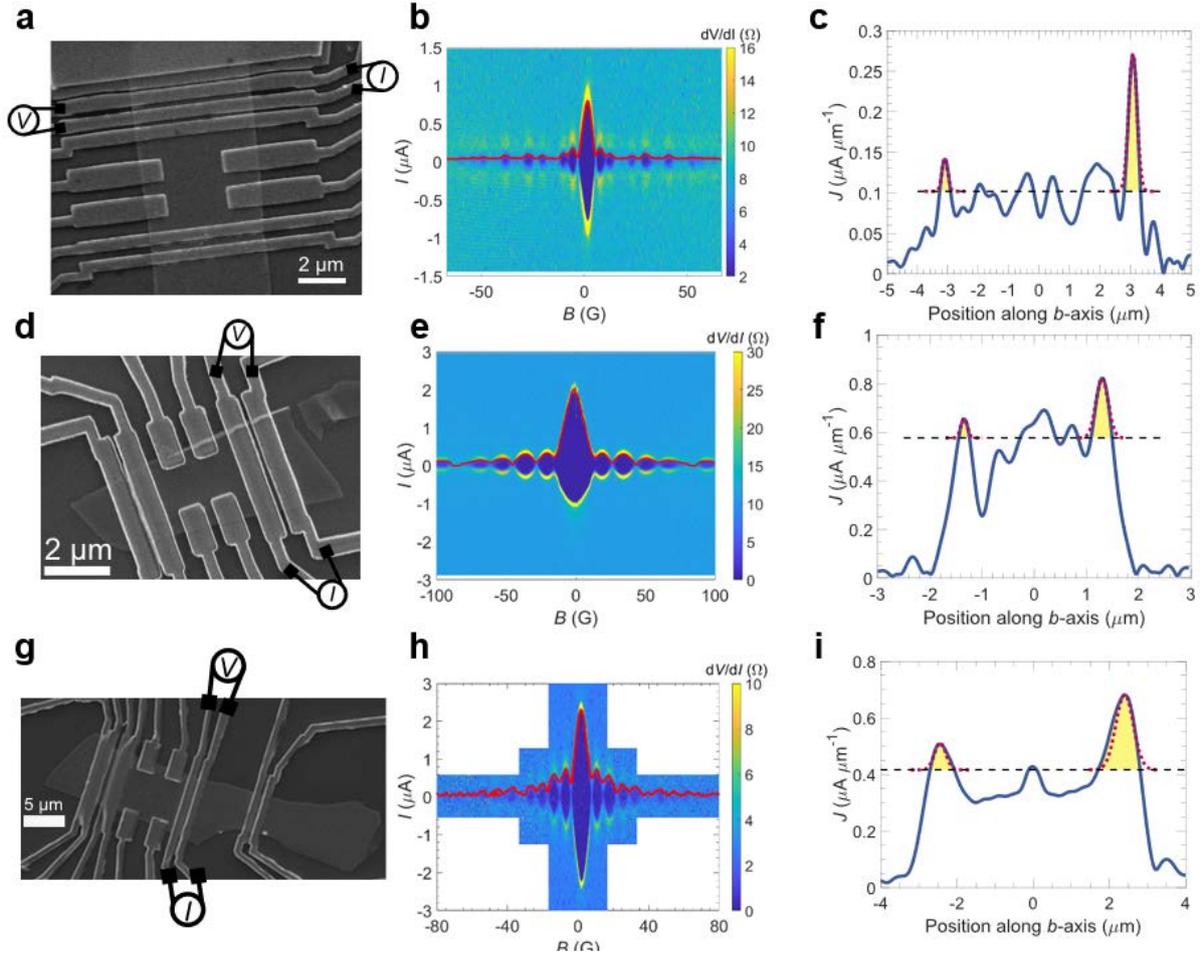

**Figure S7 | Other devices of Dev. A2, A3 and A4. a, d, g,** Scanning electron micrograph of Dev. A2, A3 and A4 with measurement configuration, respectively. **b, e, h,** Colour-coded plot of differential resistance d$V$/d$I$ as a function of bias current $I$ and perpendicular external magnetic field $B$ for Dev. A2 (**b**), Dev. A3 (**e**) and Dev. A4 (**h**). Red solid lines represent extracted Josephson critical current. **c, f, i,** Extracted spatial distribution of Josephson current density $J$ for Dev. A2 (**c**), Dev. A3 (**f**) and Dev. A4 (**i**). Dotted line represents Gaussian fitting for each edge enhancement.

## S8. Other WTe$_2$ Josephson junction devices along *b*-axis

Figure S8 show other Nb-WTe$_2$-Nb Josephson junction device of Dev. B2 (Figs. S8a and b), Dev. B3 (Figs. S8c and d), and Dev. B4 (Figs. S8e and f). Here, Dev. B2, Dev. B3, and Dev. B4 was measured under 30 mK. Information of devices are summarized in Table S1. Because of anisotropic transport, we cannot observe non-vanishing or ordinary interference pattern and critical current much smaller than Dev. A.

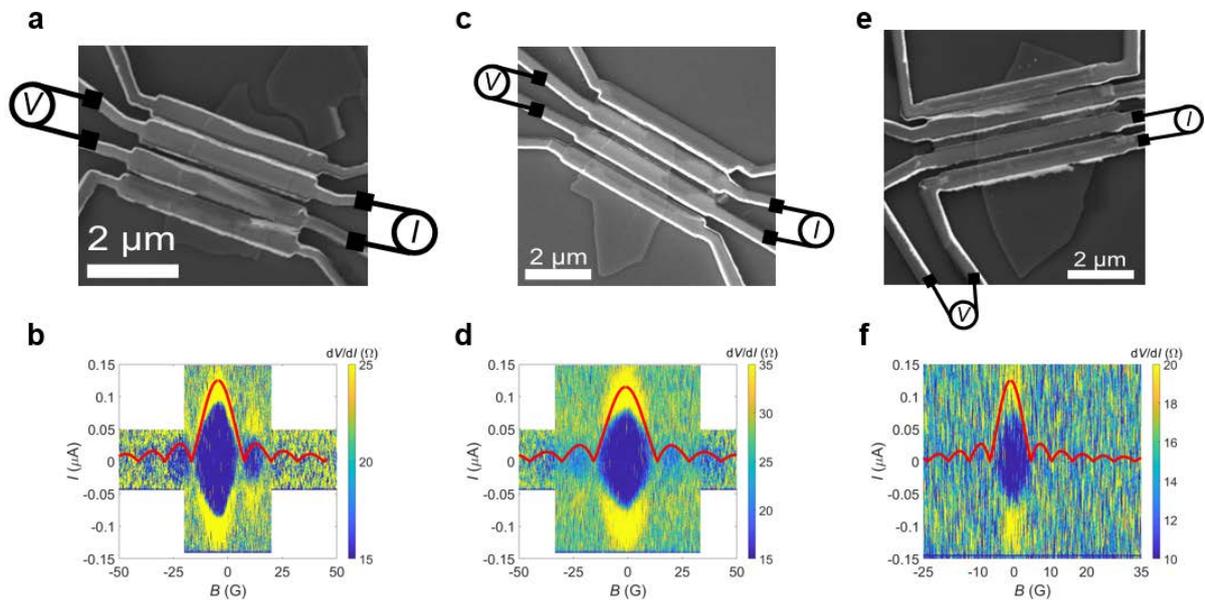

**Figure S8 | Other devices of Dev. B2, B3 and B4. a, c, e,** Scanning electron micrograph of Dev. B2, B3 and B4 with measurement configuration, respectively. **b, d, f,** Colour-coded plot of differential resistance d$V$/d$I$ as a function of bias current $I$ and perpendicular external magnetic field $B$ for Dev. B2 (**b**), Dev. B3 (**d**) and Dev. B4 (**f**).